# Possible Superconductivity at 37 K in Graphite-Sulfur Composite

**YANG Hai-Peng (杨海朋), WEN Hai-Hu（闻海虎）*, ZHAO Zhi-Wen（赵志文）,**
**LI Shi-Liang（李世亮）**
**National Laboratory for Superconductivity, Institute of Physics and Center for Condensed Matter Physics, Chinese Academy of Sciences, P. O. Box 603, Beijing 100080**

**Abstract**

Sulfur intercalated graphite composites with diamagnetic transitions at 6.7 K and 37 K are prepared. The magnetization hysteresis loops ( MHL ), X-ray diffraction patterns, and resistance were measured. From the MHL, a slight superconducting – like penetration process is observed at 15 K in low field region. The XRD shows no big difference from the mixture of graphite and sulfur indicating that the volume of the superconducting phase ( if any ) is very small. The temperature dependence of resistance shows a typical semi-conducting behavior with a saturation in low temperature region. This saturation is either induced by the de-localization of conducting electrons or by possible superconductivity in this system.

Carbon-based materials have attracted great attention due to the unique structural, magnetic and electrical properties.[1-5] It is suggested that there may exist high temperature superconductivity in materials based on carbon.[6-7] Recently the discovery of superconductivity at 117 K in fullerite $C_{60}$ causes another upsurge in this kind of material. [8] In $C_{60}$ each carbon atom has three neighbors, being quite similar to that in the graphite plane that stables the $sp^2$-valence. A naïve conjecture would be that the superconductivity exists probably in intercalated-graphite. Early studies indeed found that the superconductivity occurs in graphite intercalated by alkali-metals[9-11] although the superconducting transition temperatures are quite low. Actually the boron-sheet in the new superconductor $MgB_2$ has also a graphite-like structure.[12] It is thus tempting to seek the high temperature superconductivity in intercalated graphite.

To prepare C-S composites, both graphite powder and sulfur powder were mixed in weight ratio of C : S = 1 : 1. The mixture was pressed into pellets. The pellets were wrapped with Ta foil and were sealed in vacuumized quartz tube. Then the tube was heated at 400℃ for several hours and then cooled to 150℃ and followed by remaining at this temperature for ten hours. We prepared samples with other nominal weight ratios ( C : S = 1 : 3, 1 : 2, 2 : 1, 3 : 1, 8 : 1 ). It is found that the samples with C / S weight ratios between 1 : 1 and 3 : 1 have diamagnetic transition at about 37 K. In some samples, the onset transition temperature is 32 K. The C : S = 1 : 1 sample has the strongest diamagnetic signal. Therefore, the following measurements are all based on the samples with C / S = 1 : 1. The diamagnetic properties were measured by using a superconducting quantum interference device (Quantum Design SQUID, MPMS 5.5 T). The resistance was measured by using the standard four-probe technique.

Figure 1 shows two diamagnetic transitions on the C : S = 1 : 1 sample: one is at about 6.7 K and the other is at about 37 K. The external magnet field used here is 10 Oe. One week later the $M(T)$ curve was re-measured and only the transition step at 37 K was observed, the transition at about 6.7 K was disappeared. One month later both transitions were disappeared. This effect may be induced by the fact that the intercalated sulfur will escape from the structure. In order to know whether the diamagnetic step at 37 K is induced by a superconducting transition, we measured the magnet hysteresis loop (MHL) at 5 K and 15 K. Figures 2(a), 2(b) and 2(c) show the MHL results. At 5K and 15K, the sample clearly exhibits a hysteresis loop. Interestingly, the MHL near zero field looks very much like that of a type-II superconductor: The penetration process of magnetic flux gives rise to a up-turn from the initial Meissner slope (as shown by the dashed line).

A real superconducting transition should be characterized by two properties: zero resistance and diamagnetic moment. These are induced by the phase coherence of the Cooper pairs.

In Fig.3 we show the resistive $R(T)$ curve of the same sample. Figure 3(a) shows the raw data. The overall shape of the $R(T)$ curve is semiconductor-like besides a saturation in low temperature region. There are two possibilities for the low temperature saturation of resistance: de-localization of conduction electrons and superconductivity. Assuming that the superconductivity occurs really in low temperature region, since the superconducting volume is very small (below 0.1% as seen from the diamagnetic signal), the superconducting islands are berried in the semiconducting graphite matrix. Therefore, the resistance contains a background from the graphite and a slight down-ward turn in the low temperature region due to the formation of isolated superconducting regions. According to this picture one can make a polynomial fit to the high temperature data to obtain the background signal. Subtraction from the total signal with this background signal will amplify the superconducting transition. The background signal is shown in Fig.3(a) by the dashed line and the subtracted signal is shown in Fig.3(b). Interestingly a superconducting-like transition occurs at about 70 K and a sharper transition occurs at about 37 K being consistent with the diamagnetic measurement. While it is important to note that the data after mathematical treatment as shown in Fig.3(b) hinges on the occurrence of superconductivity rather than giving a direct evidence of superconductivity. To have a clear resistive evidence for superconductivity, one needs to increase the superconducting volume and see a clear dropping-step of resistance.

The mixture of raw materials of graphite and sulfur has also been measured with the SQUID. No diamagnetic transition on $M(T)$ curve has been observed. Thus the diamagnetic transition in the reacted C/S composite is resulted from the properties of the material. In order to see any structural change before and after the chemical reaction we have performed the XRD measurement. From the literature we know that the reaction temperature is not high enough to destroy the layered structure of graphite. The spacing between two graphite layers is 3.35 Å and the atom radius of sulfur is 1.03 Å, thus the sulfur can diffuse into the spacing between the graphite layers. In addition, the electronegativity of graphite and sulfur are 2.54 and 2.58, respectively. The difference of their electronegativity is so small that it is hard to say that carbon or sulfur can give or accept electrons from each other. In this case the electrons from carbon and sulfur will probably form some kinds of covalence bonding. The intercalated sulfur will make the graphite plane to curve slightly leading to a change to the electron band structure of graphite. A series of XRD data with the content of sulfur from 75 % (C / S = 1 : 3) to 11.1 % (C / S = 8 : 1) are shown in Fig. 4 (from bottom to top, all data add 5 in turn). From the XRD curves we cannot find large difference between the reacted composite and the initial mixture. These may suggest that the superconducting phase (if any) has only a very small volume and the major part of the graphite has not been affected by the sulfur intercalation.

Recently Ricardo da Silva *et al.*[13] reported the similar property in the C/S composite. Together with our results we conclude that a diamagnetic transition at 37 K surely happens in sulfur intercalated graphite. The initial part of the MHL also shows the behaviour like a type-II superconductor. The evidence from the resistive transition is still lacking although saturation of the $R(T)$ curve has been observed in the low temperature region. In analogy to the structure and electronic property of $MgB_2$ and $C_{60}$, it is suggested that an instable superconductivity may have occurred in sulfur intercalated graphite.

**Ackowledgement**

Supported by the National Natural Science Foundation of China under Grant No. 19825111, and the Ministry of Science and Technology of China (NKBRSF-G19990646).

* hhwen@aphy.iphy.ac.cn

## Figure Captions

**Fig.1** Temperature dependence of magnetization of the sulfur intercalated graphite. A clear diamagnetic transition occurs at 37 K and 6.5 K respectively.

**Fig.2** Magnetization hysteresis loops at (a) 15 K; (b) an enlarged view at 15 K; and (c) 5 K. From Fig.2(b) and Fig.2(c) one can see a superconducting-like magnetic penetration process. The dashed line represents the Meissner-line.

**Fig.3** Temperature dependence of (a) resistance and (b) the subtracted resistance. The dashed line in Fig.3(a) shows a polynomial fit ( $R_{fit}$ ) to the experimental data between 80 K and 295 K.

**Fig.4** XRD patterns for pure graphite, sulfur, the mixture of graphite and sulfur, sample1-6 corresponding to content of sulfur from 75% to 11.1 %. It is clear that the structure of the composites before and after the reaction has no large difference indicating the small volume of the superconducting phase ( if any ).

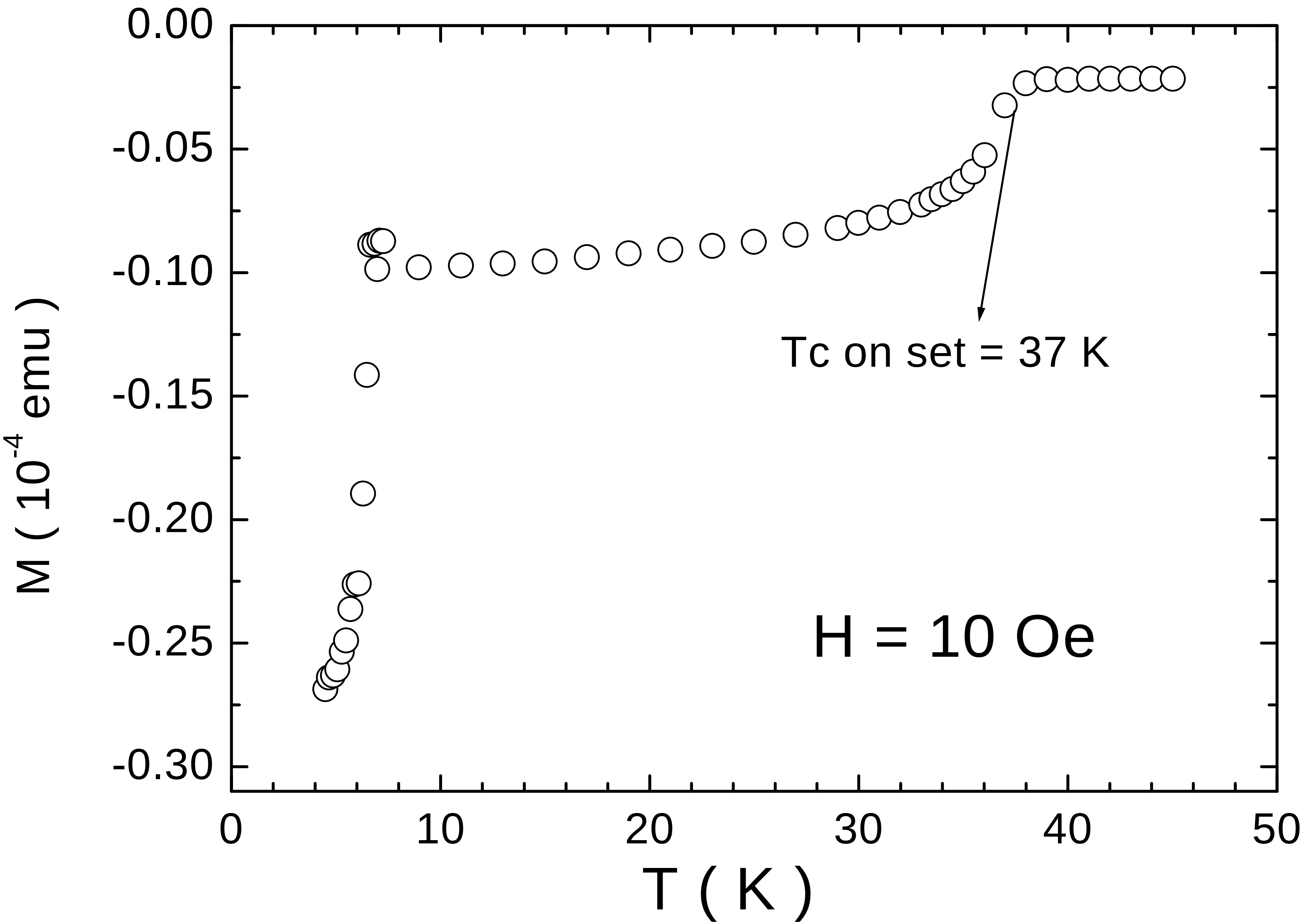

0.00
-0.05
-0.10
-0.15
-0.20
-0.25
-0.30
M ( 10^-4 emu )
0
10
20
30
40
50
T ( K )
Tc on set = 37 K
H = 10 Oe

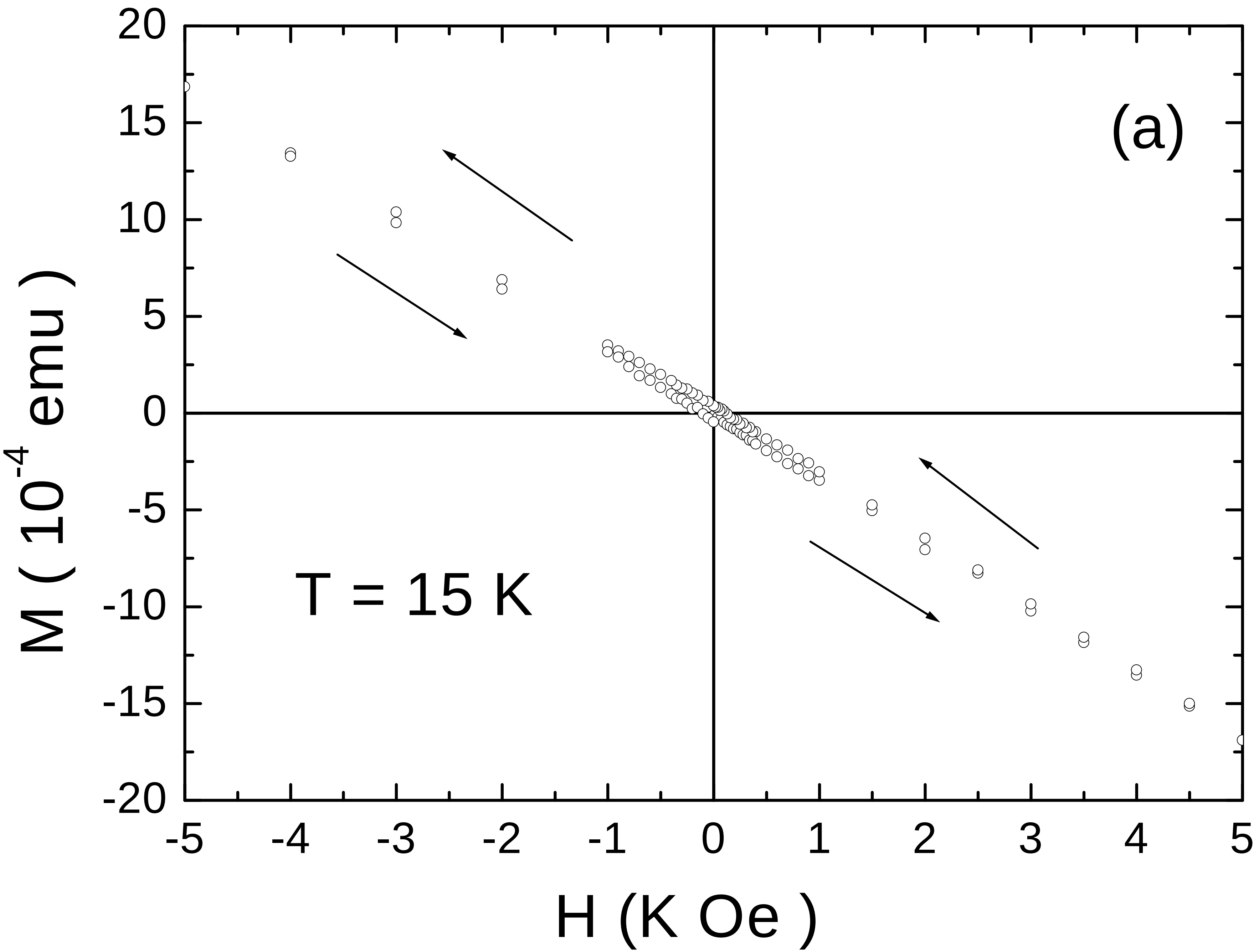

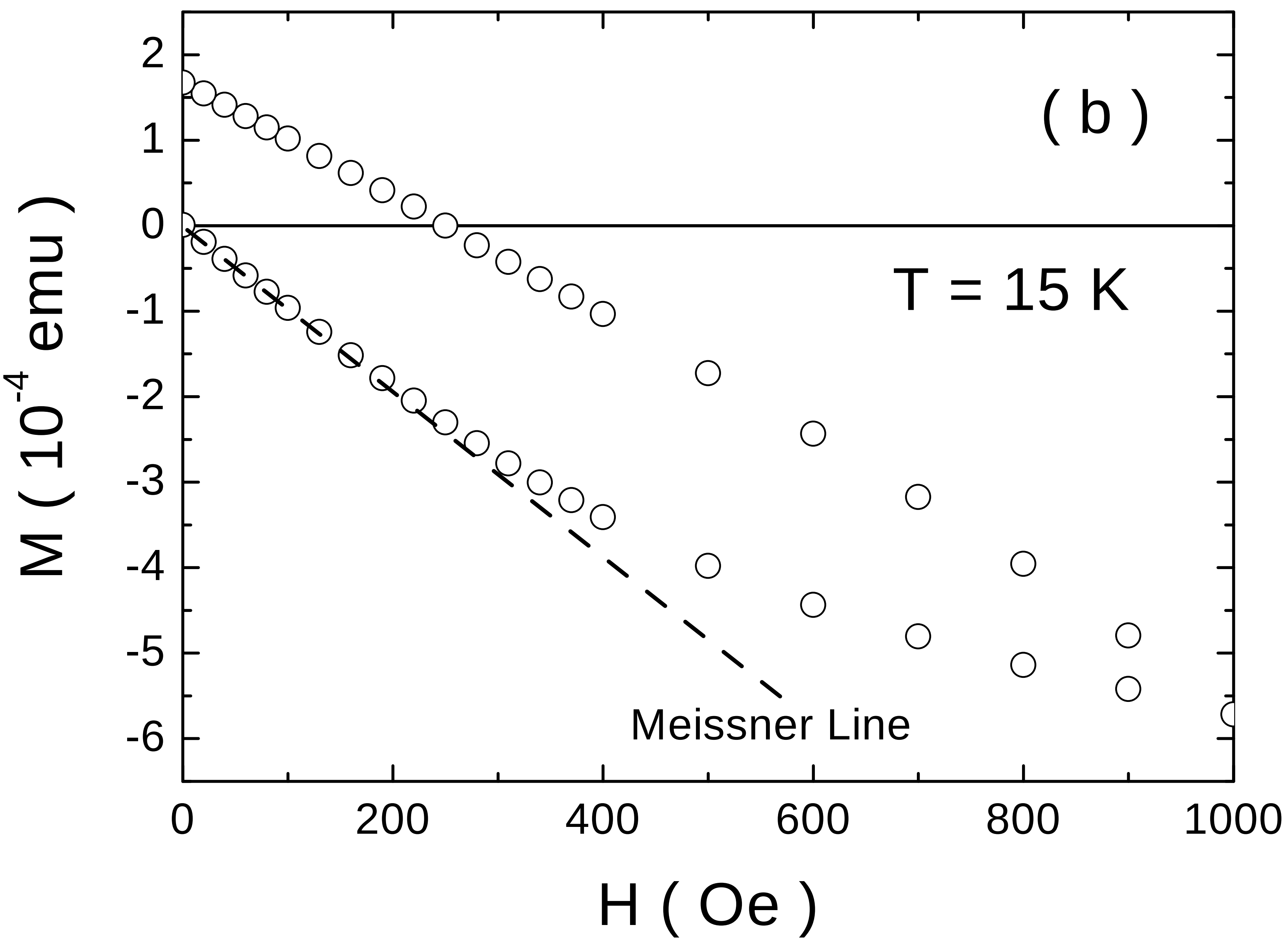

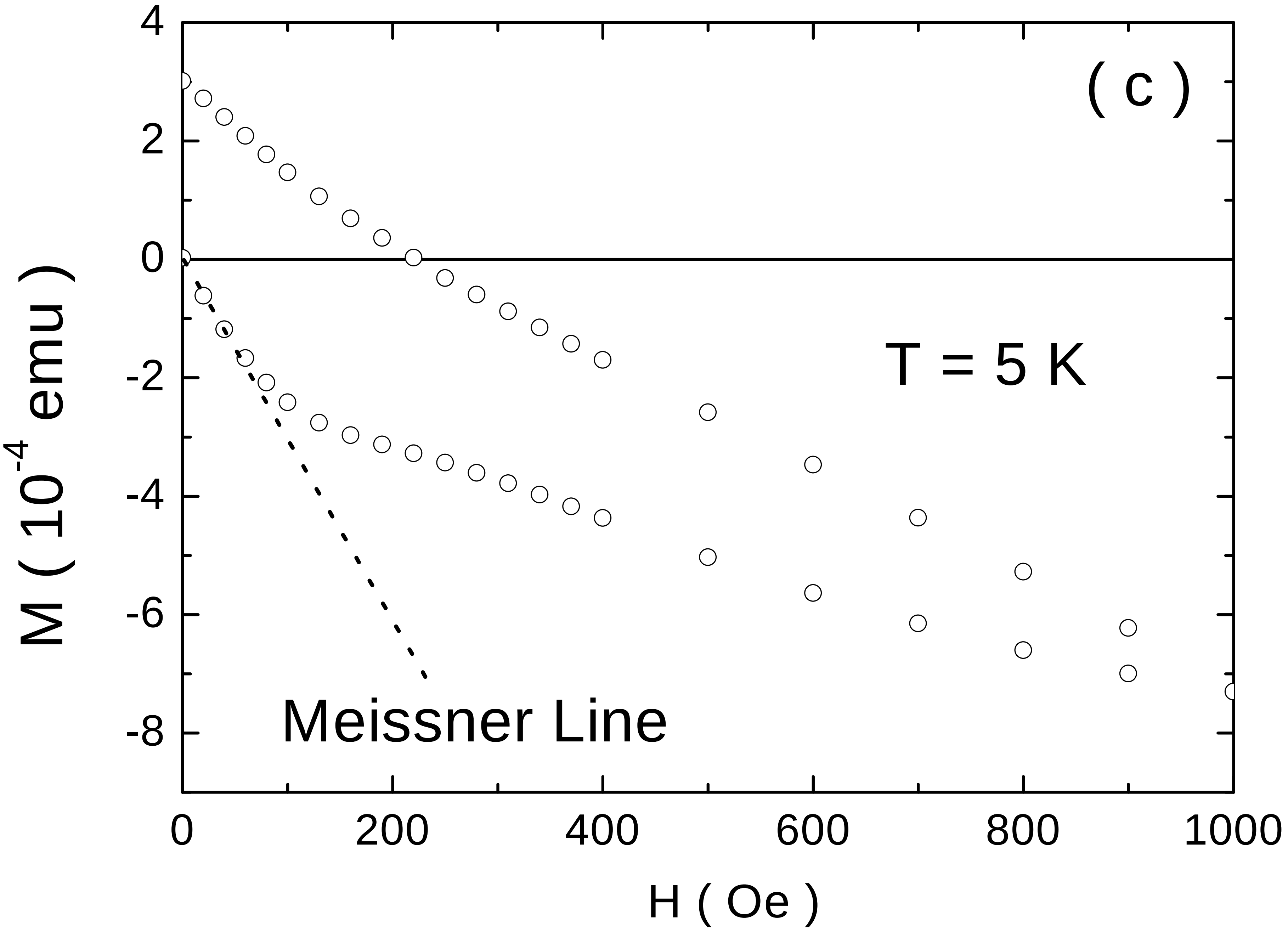

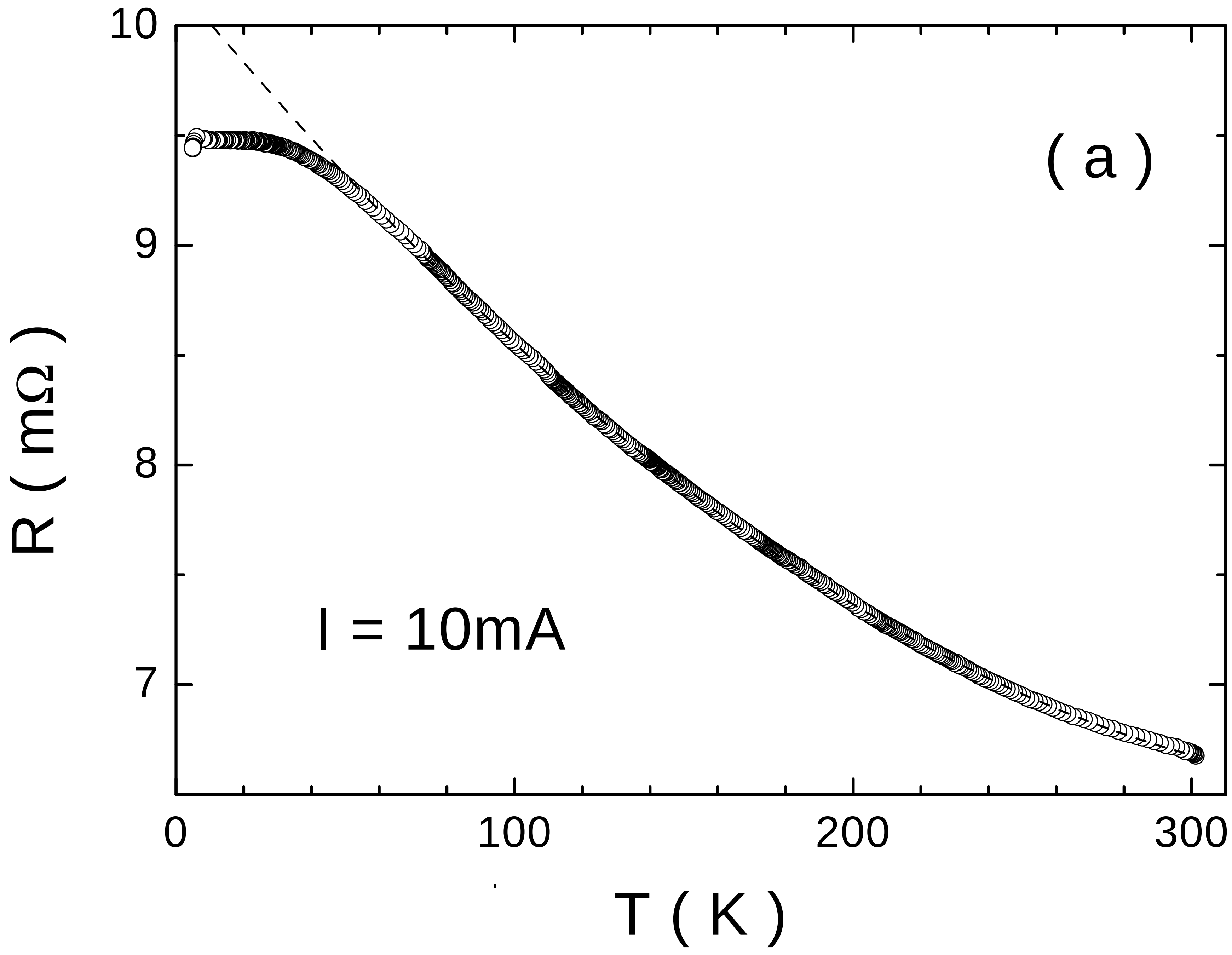

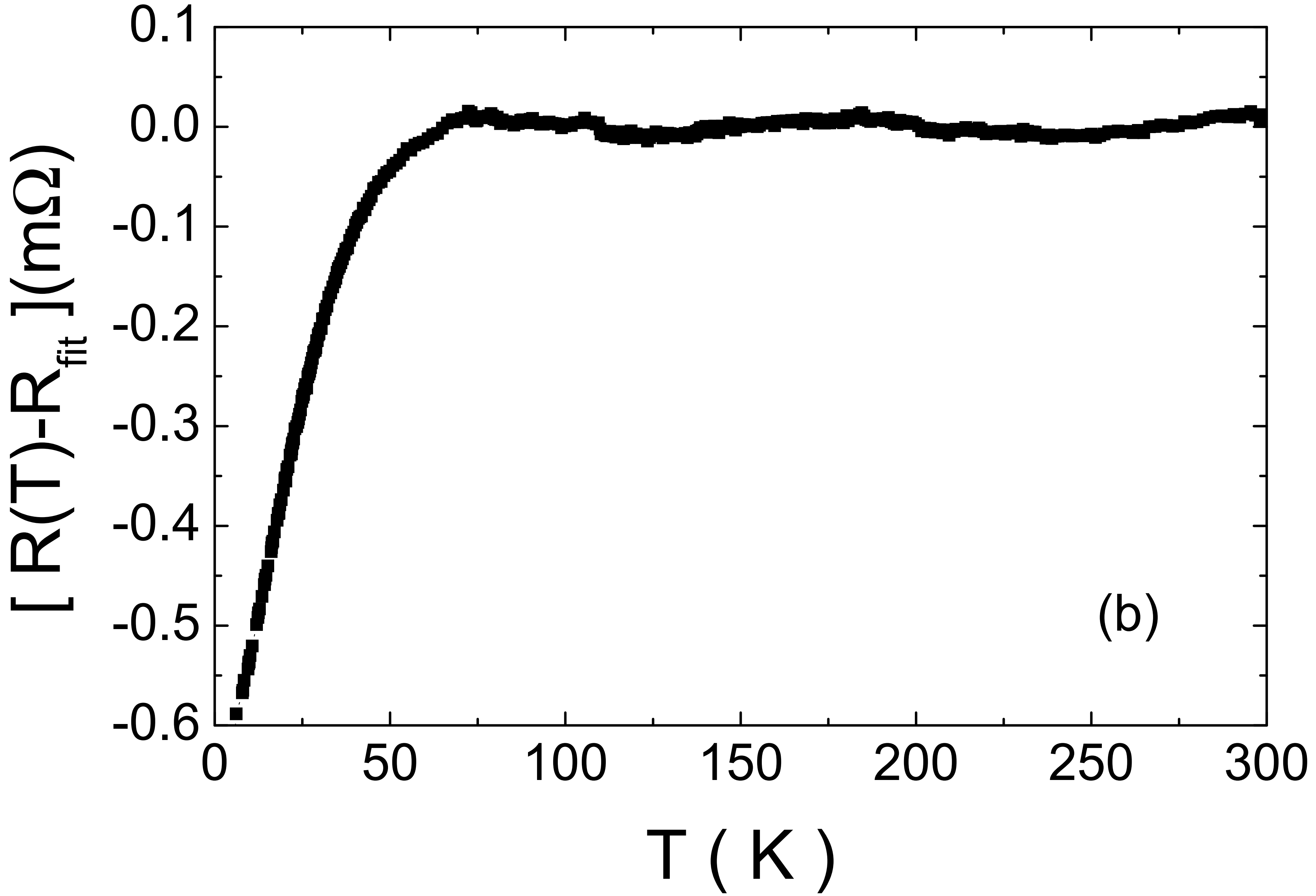

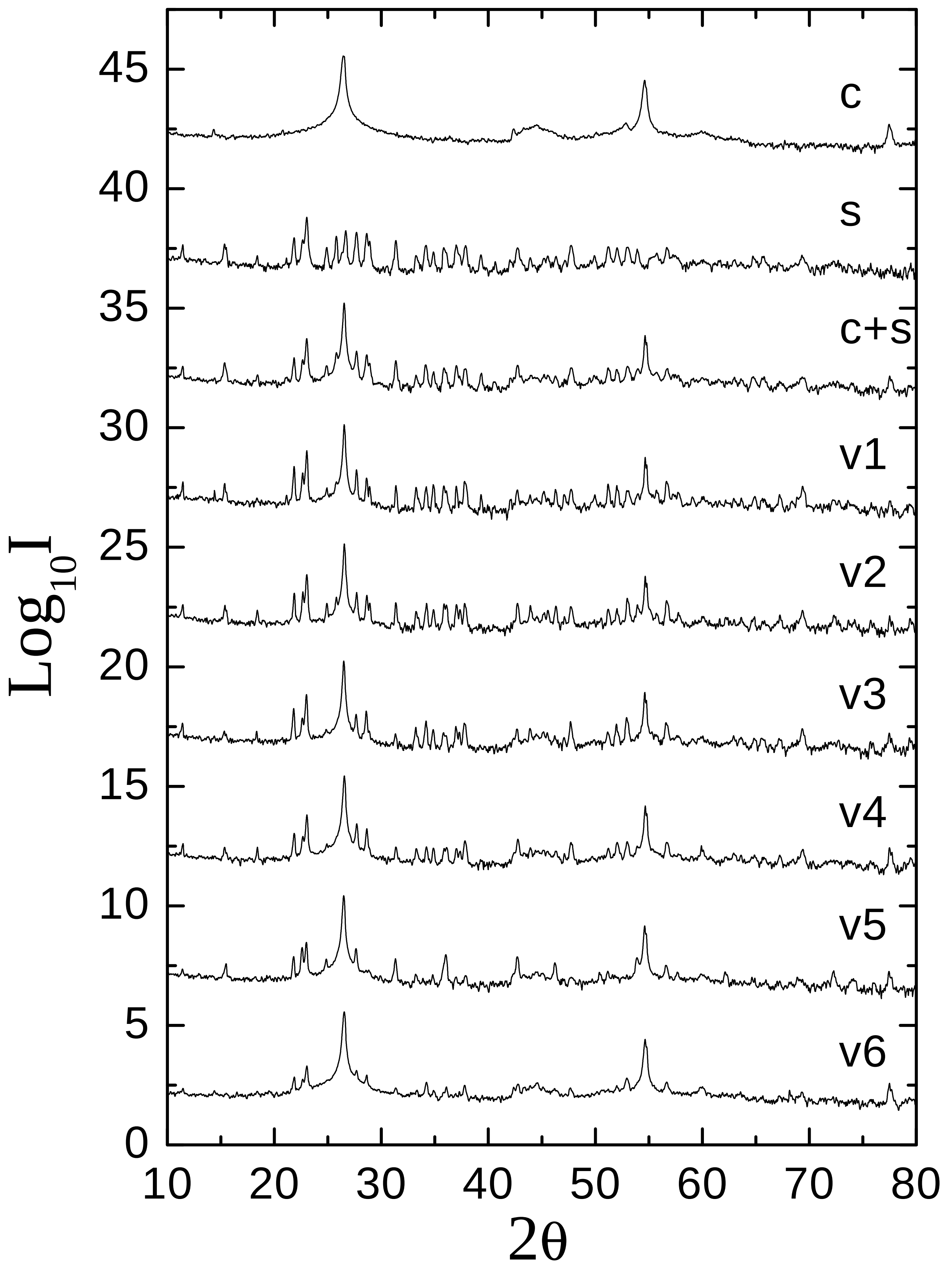